\pdfoutput=1

\documentclass[9pt,twocolumn]{article}

\usepackage{authblk}
\usepackage{acronym}
\usepackage{amsmath}
\usepackage{cleveref}
\usepackage{graphicx}
\usepackage{siunitx}

\newacro{CFL}{cell-free layer}
\newacro{DPD}{dissipative particle dynamics}
\newacro{RBC}{red blood cell}
\newacro{SFS}{stress-free state}
\newacro{TTF}{tank-treading frequency}

\newcommand{\Ht}{\mathrm{Hct}}
\DeclareSIUnit{\cmHHO}{\text{cmH}_2\text{O}}

\newcommand{\key}[1]{\raisebox{0.0pt}{\protect\includegraphics{figures/keys/#1}}}

\title{The Volume of Healthy Red Blood Cells is Optimal for Advective Oxygen Transport in Arterioles}

\author[a,c]{Lucas Amoudruz}
\author[b]{Athena Economides}
\author[a,*]{Petros Koumoutsakos}

\affil[a]{Computational Science and Engineering Laboratory, School of Engineering and Applied Sciences, Harvard University, Cambridge, MA 02138, United States}
\affil[b]{Institute of Neuropathology, University of Zurich, CH-8091 Zurich, Switzerland}
\affil[c]{Computational Science and Engineering Laboratory, ETH Z\"{u}rich, CH-8092, Switzerland}
\affil[*]{Corresponding author: petros@seas.harvard.edu}

\date{}

\begin{document}
\twocolumn[
  \begin{@twocolumnfalse}
    \maketitle
    \begin{abstract}
  Red blood cells (RBCs) are vital for transporting oxygen from the lungs to the body's tissues through the intricate circulatory system.
  They achieve this by binding and releasing oxygen molecules to the abundant hemoglobin within their cytosol.
  The volume of RBCs affects the amount of oxygen they can carry, yet whether this volume is optimal for transporting oxygen through the circulatory system remains an open question.
  This study explores, through high-fidelity numerical simulations, the impact of RBC volume on advectve oxygen transport efficiency through arterioles which form the area of greatest flow resistance in the circulatory system.
  The results show that, strikingly, RBCs with volumes similar to those found in vivo are most efficient to transport oxygen through arterioles.
  The flow resistance is related to the cell-free layer thickness, which is influenced by the shape and the motion of the RBCs: at low volumes RBCs deform and fold while at high volumes RBCs collide and follow more diffuse trajectories.
  In contrast, RBCs with a healthy volume maximize the cell-free layer thickness, resulting in a more efficient advectve transport of oxygen.
    \end{abstract}
  \end{@twocolumnfalse}
]


Blood flow plays a vital role in sustaining life as it enables the delivery of oxygen and nutrients to every tissue and organ in the body while also removing harmful waste like carbon dioxide.
\Acp{RBC} are a crucial component of blood, making up almost half of the total blood volume in humans~\cite{popel1989}. These specialized cells are responsible for carrying oxygen from the lungs to the rest of the body, where it is used to fuel cellular metabolism.
They are formed by a visco-elastic membrane that surrounds the cytosol of the cell.
The cytosol contains a high concentration of hemoglobin, which is responsible for the transportation of oxygen~\cite{popel1989,di2015}.
The \ac{RBC} membrane is composed of a lipid bilayer anchored to a cytoskeleton and allows for significant deformations~\cite{lim2008}.
The lipid bilayer of the membrane behaves like an incompressible liquid crystal with localized bending resistance, while the cytoskeleton provides resistance to local shear and dilation~\cite{geekiyanage2019}.
According to current models, the unstressed shape of the cytoskeleton is an oblate spheroid of reduced volume approximately $0.95$, compared to a sphere with the same membrane area~\cite{lim2008, geekiyanage2019, amoudruz2023stress}.
However, at equilibrium healthy \acp{RBC} adopt a biconcave shape with a reduced volume ($v$) of approximately $0.65$~\cite{evans1972a}.
In this paper, we study the effect of this reduced volume on the oxygen transport efficiency in straight tubes with circular cross-section that have representative characteristics of arterioles.

The biconcave shape and reduced volume of \acp{RBC} as well as  metrics for their optimality remain a subject of debate in the scientific community.
Several researchers proposed that the biconcave shape allows for a large surface-to-volume ratio for  \acp{RBC} so that they maximize oxygen exchange between the \ac{RBC} interior and their environment~\cite{lenard1974, uzoigwe2006}.
The exchange of oxygen between the cytoplasm and the environment is higher for cells with lower thickness because cytoplasmic diffusion may be more limiting than membrane permeability, as shown by Richardson et al.~\cite{richardson2020}.
Other studies suggest that the biconcave shape allows for a more efficient flow through capillaries and kidney tubules, due to the \acp{RBC}' deformability~\cite{lim2008, pivkin2016, vahidkhah2016, namvar2021}.
Uzoigwe~\cite{uzoigwe2006} proposed that \acp{RBC} have a biconcave shape to maximize their moment of inertia, thereby reducing shear stresses in blood flow and decreasing blood flow resistance in arteries, but did not present any quantitative results.
Guo et al.~\cite{guo2017} showed that the volume of cells affects their stiffness, while other studies found that osmotic pressure and cell stiffness affect the viscosity of whole blood~\cite{strumia1963,stone1968,reinhart1992,schmid2010}.
Alterations in \ac{RBC} volume can significantly impact the function of the circulatory system and individual health~\cite{uzoigwe2006}.
Farutin et al.~\cite{farutin2018} studied numerically the cell transport efficiency in channels and observed that the optimal hematocrit depends on the \acp{RBC}' volume.

In this study, we investigate the effect of the reduced volume of \acp{RBC} on the advectve oxygen transport efficiency along the blood flow direction.
To this end, we study the transport of a fixed quantity of hemoglobin by \acp{RBC} with various resting shapes, parameterized by the reduced volume of the cells.
We investigate the effect of the \ac{RBC} reduced volume on the oxygen flux through a straight tube with circular cross-section.
This metric has been used experimentally to determine the optimal hematocrit for oxygen transport~\cite{stone1968,farutin2018}.
We choose tubes with dimensions and pressure gradient similar to those found in arterioles, where the flow resistance is the highest in the circulatory system~\cite{bohlen1977, chilian1986, meininger1987, mulvany1990}.

The study relies on a recent and appropriately validated \ac{RBC} model, comprising visco-elastic membranes enclosing the cytosol and suspended in the blood plasma.
The model was extensively calibrated in Amoudruz et al.~\cite{amoudruz2023stress} and  validated against experimental data in various flow conditions.
The evolution of the solvent and the cytosol are described by \ac{DPD}~\cite{hoogerbrugge1992,espanol1995}.
The numerical simulations are performed with \textit{Mirheo}, a high-performance software for blood flow and microfluidics~\cite{alexeev2020a}.

\section*{Methods}

We model blood with \acp{RBC} composed of visco-elastic membranes surrounding their viscous cytosol, and suspended in the blood plasma.
The \ac{RBC} membrane deforms from forces that arise from bending resistance of the lipid-bilayer,
as well as due to the shear and dilation elasticity of the cytoskeleton with respect to its \ac{SFS}, and membrane viscosity.
The resistance to bending is described by the energy
\begin{equation} \label{eq:energy:bending}
  U_{bending} = 2 \kappa_b \oint{H^2dA},
\end{equation}
where the integral is taken over the membrane surface, $\kappa_b$ is the bending modulus and $H$ is the local mean curvature.
The in-plane elastic energy is given by
\begin{equation*}
  \begin{split}
    U_{in-plane} =& \frac {K_\alpha}{2} \oint {\left( \alpha^2 + a_3 \alpha^3 + a_4 \alpha^4 \right) dA_0} \\
    &+ \mu \oint{ \left( \beta + b_1 \alpha \beta + b_2 \beta^2 \right) dA_0},
  \end{split}
\end{equation*}
where the integral is taken over the \ac{SFS} surface, $\alpha$ and $\beta$ are the local dilation and shear strain invariants, respectively, $K_\alpha$ is the dilation elastic modulus, $\mu$ is the shear elastic modulus and the coefficients $a_3$, $a_4$, $b_1$ and $b_2$ are parameters that control the non-linearity of the membrane elasticity for large deformations~\cite{lim2002}.

Each membrane is composed of 2562 particles located at the corners of a triangulated mesh and evolve according to Newton's law of motion.
The bending energy is discretized following references~\cite{julicher1996,bian2020a}, and the in-plane energy is computed as described in Lim et al.~\cite{lim2008}.
The forces acting on the particles are the negative gradients of the discretized energy terms, with respect to the particle positions.
Dissipation on the membrane is modeled by  particles sharing an edge in the triangle mesh that exert a pairwise force as described in Fedosov~\cite{fedosov2010pHD}, proportional to the membrane viscosity $\eta_m$.
Finally, the area of the membrane and the volume of the cytosol are constrained through energy penalization terms,
\begin{equation*}
  U_{area} = k_A \frac{\left(A - A_0\right)^2}{A_0}, \quad
  U_{volume} = k_V \frac{\left(V - V_0\right)^2}{V_0},
\end{equation*}
where $A_0$ and $V_0$ are the area and volume of the cell at rest and $A$ and $V$ are the area and volume of the cell, respectively.
These penalization terms are required as the \ac{DPD} method does not impose incompressibility of the cytosol, and the membrane elastic energies alone do not conserve the membrane area.
The parameters of the model take physiological values calibrated from experimental data~\cite{amoudruz2023stress} and are listed in the Appendix.

The \ac{RBC} cytosol and surrounding plasma are represented with particles that evolve through \ac{DPD} interactions~\cite{hoogerbrugge1992,espanol1995,amoudruz2023stress} (see Appendix for more details).
We emphasize that the model parameters correspond to the 1:5 ratio of viscosities for the plasma ($\SI{1.2}{\pascal\second}$~\cite{wells1961shear}) and the \ac{RBC} cytosol ($\SI{6}{\pascal\second}$~\cite{wells1969red}).
To model the no-slip and no-flux boundary conditions on walls, particles are bounced-back from the walls' surface.
Furthermore, the particles interact with a layer of frozen particles that are inside the walls through \ac{DPD} interactions~\cite{revenga1998}.
In addition, \ac{DPD} particles are bounced-back from membrane surfaces, and interact with the membrane particles through the dissipative and stochastic parts of the \ac{DPD} interactions only~\cite{fedosov2010pHD}.
The cytosol and plasma particles interact with each other only through the conservative part of the \ac{DPD} forces.
The \ac{DPD} parameters are chosen from the macroscopic properties of the fluids as described in ref.~\cite{amoudruz2022thesis}.

\section*{Results}

\begin{figure}[h]
  \centering
  \includegraphics{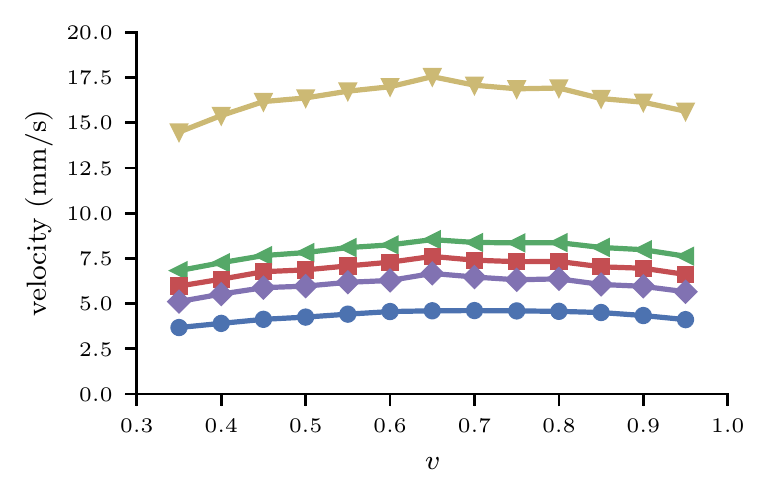}
  \caption{Mean flow velocity in the tube against the reduced volume of the RBCs for a fixed pressure gradient, with different tube radii, hematocrits and pressure gradients.\\
    \key{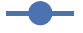}: $R=\SI{30}{\micro\meter}$, $\Ht=40\%$, $\nabla p = \SI{1e-3}{\cmHHO\per\micro\meter}$;
    \key{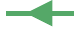}: $R=\SI{40}{\micro\meter}$, $\Ht=40\%$, $\nabla p = \SI{1e-3}{\cmHHO\per\micro\meter}$;
    \key{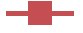}: $R=\SI{40}{\micro\meter}$, $\Ht=45\%$, $\nabla p = \SI{1e-3}{\cmHHO\per\micro\meter}$;
    \key{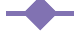}: $R=\SI{40}{\micro\meter}$, $\Ht=50\%$, $\nabla p = \SI{1e-3}{\cmHHO\per\micro\meter}$;
    \key{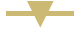}: $R=\SI{40}{\micro\meter}$, $\Ht=45\%$, $\nabla p = \SI{2e-3}{\cmHHO\per\micro\meter}$.
    \label{fig:blood:velocity}
  }
\end{figure}

\begin{figure*}
  \centering
  \includegraphics[width=\textwidth]{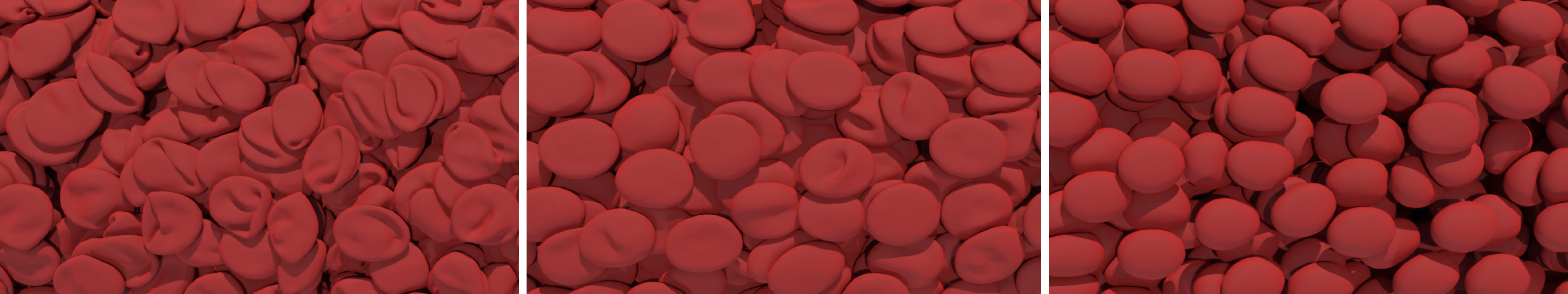}
  \caption{Snapshots of RBCs flowing near the walls of the tube with $R=\SI{40}{\micro\meter}$ and $\Ht=0.45$.
    The flow direction is from left to right.
    Cells with different reduced volumes respond differently to the flow shear (from left to right: $v=0.35$, $0.65$, $0.95$).
    \label{fig:snapshots}
  }
\end{figure*}

We examine the equilibrated flow of \acp{RBC} inside a circular tube of radius $R \in \{\SI{30}{\micro\meter}, \SI{40}{\micro\meter}\}$.
These sizes are typical for arterioles~\cite{fung1998}.
The volume fraction of \acp{RBC}, or tube hematocrit, is set to $\Ht\in \{40\%,45\%,50\%\}$ and the blood suspension is subjected to a pressure gradient $\nabla p \in \{\SI{1e-3}{\cmHHO\per\micro\meter}, \SI{2e-3}{\cmHHO\per\micro\meter}\}$, typical for arterioles of this size~\cite{fung1998}.
These conditions correspond to mean flow velocities of the order of $\SI{10}{\milli\meter\per\second}$, consistent with values reported in the literature~\cite{popel1989}.
Mean flow velocities of all simulations are reported on \cref{fig:blood:velocity}.
We study different cases where all \ac{RBC} membranes have the same area and visco-elastic properties (calibrated from experimental data~\cite{amoudruz2023stress}) but have a different volume $V_0$ parameterized by the reduced volume $v = V_0/V_s$, where $V_s$ is the volume of a sphere with the same area as the cell's membrane.
All initial triangle mesh were produced by minimizing the membrane energy with a target volume $V_0$.
The number of \acp{RBC} is adapted to keep the hematocrit constant, $N = \lfloor \pi R^2 L \Ht / V_0 \rfloor$.
We study a periodic domain with a length $L = \SI{100}{\micro\meter}$ along the flow direction.
No-slip and no-flux boundary conditions are applied on the side walls.
All quantities are reported from equilibrated blood flows.

We find that the reduced volume affects the flow patterns and \acp{RBC} deformability (\cref{fig:snapshots}).
Specifically,  for low reduced volumes, the cells deform much more than at high values of $v$.
In the limit $v\rightarrow 1$, only spherical cells are allowed, and as the area is constant they cannot deform.
\Acp{RBC} with a physiological reduced volume do not exhibit significant deformations compared to those with lower reduced volumes.
Instead, \acp{RBC} with $v=0.65$ seem to keep a relatively flat shape similar to tank-treading \acp{RBC}, a type of motion where the membrane rotates around a steady elongated shape similar to tank treads~\cite{fischer1980}.

\begin{figure}
  \centering
  \includegraphics[width=\columnwidth]{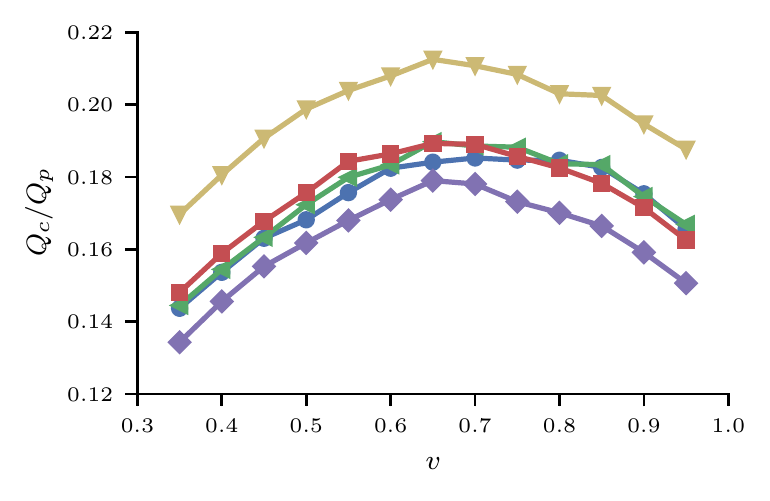}
  \caption{Cytosol flux $Q_c$, normalized by that of the plasma at zero hematocrit, $Q_p$, against the reduced volume of RBCs $v$ flowing in the tube of different radii and hematocrits.
    Same labels as in \cref{fig:blood:velocity}.
    \label{fig:vQ}
  }
\end{figure}

We quantify the amount of transported oxygen by computing  the flux of cytosol in the tube for each value of $v$,
\begin{equation*}
    Q_c = \frac{V_0}{L} \sum\limits_{i=1}^{N} U_i,
\end{equation*}
where $U_i$ is the time-averaged velocity of the $i^\text{th}$ \ac{RBC}'s center of mass.
We assume that the hemoglobin concentration inside the cytosol is constant, and therefore the cytosol flux is proportional to the oxygen flux.
\Cref{fig:vQ} shows the flux of cytosol against the reduced volume of the \acp{RBC} for different tube radii and hematocrits.
The flux of cytosol is normalized by the flux of plasma with no \acp{RBC}, $Q_p = |\nabla p| \pi R^4 / 8 \eta$, where $\eta$ is the dynamic viscosity of the plasma.
In these conditions, $Q_c$ reaches a maximum at a reduced volume $v\approx 0.65$.
Strikingly, the oxygen flux is maximized at the physiological reduced volume of \acp{RBC}.

\begin{figure}
  \centering
  \includegraphics[width=\columnwidth]{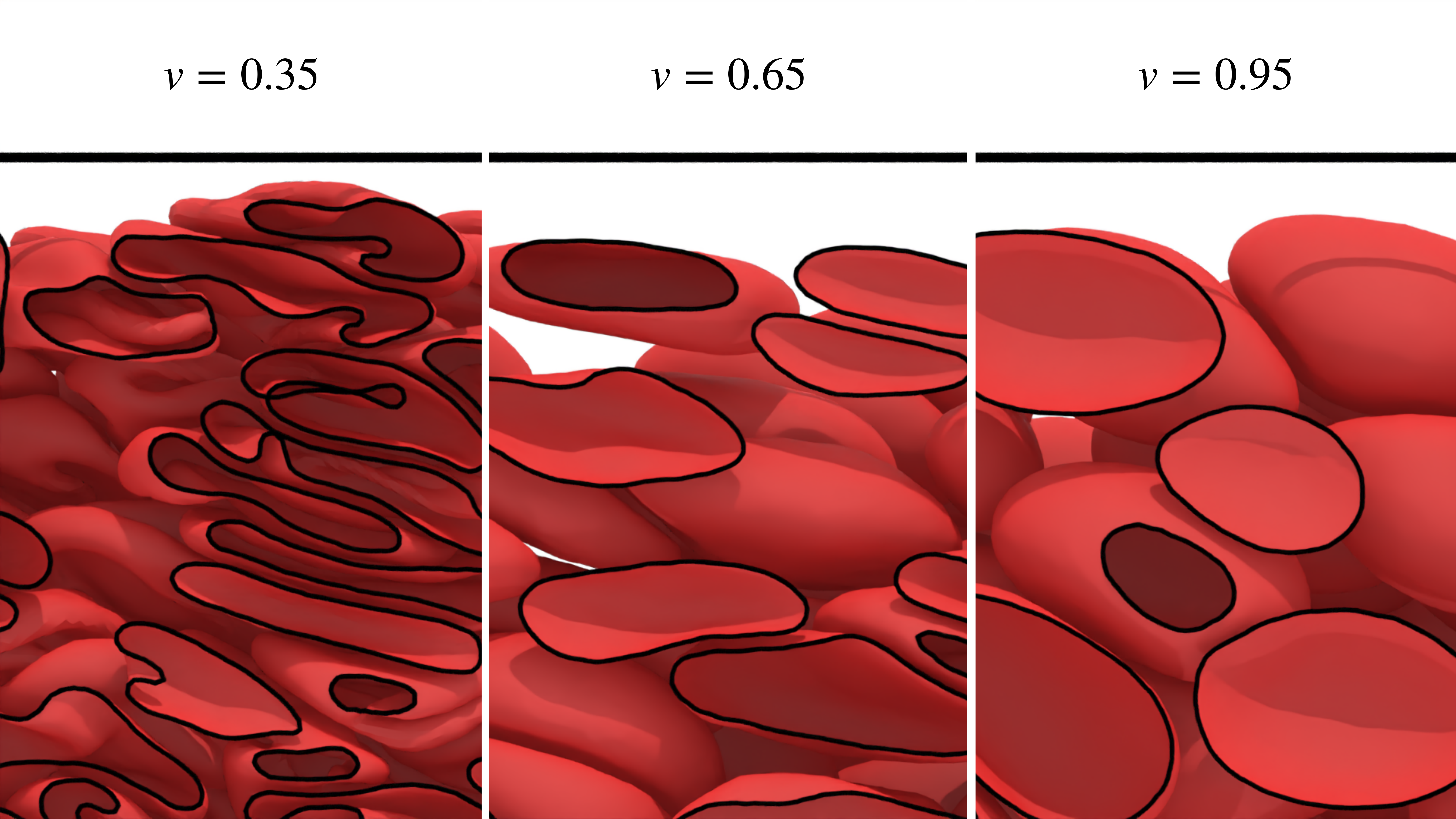}
  \caption{Snapshots of RBCs flowing near the walls (top horizontal line) of the tube. The flow direction is from left to right.
    The slice of the RBCs in the vertical plane is represented with black lines.
    \label{fig:zoom_CFL}
  }
\end{figure}

\begin{figure}
  \centering
  \includegraphics[width=\columnwidth]{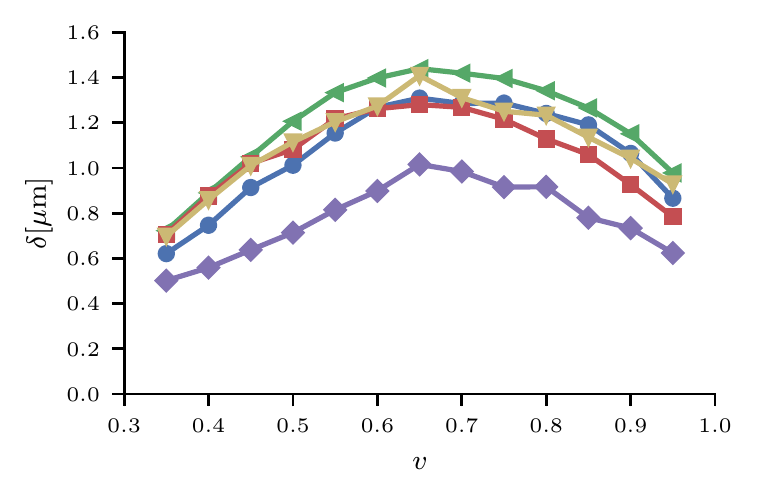}
  \caption{Cell-free layer thickness $\delta$ against the reduced volume of RBCs $v$ flowing in the tube.
    Same labels as in \cref{fig:blood:velocity}.
    \label{fig:vcfl}
  }
\end{figure}

To elucidate the mechanisms that reduce the cytosol flux when the volume of the \acp{RBC} deviates from its physiological value, we examine the \ac{CFL} near the walls of the tube (\cref{fig:zoom_CFL}).
We report its thickness, $\delta$, with respect to the reduced volume $v$ on \cref{fig:vcfl}.
The viscosity is lower in the \ac{CFL} than in the blood suspension, hence the shear rate is higher in the \ac{CFL} region.
Therefore, for a given pressure difference, a larger \ac{CFL} thickness causes a higher flux of hemoglobin (\cref{fig:vQ,fig:vcfl})~\cite{cokelet1991}.
We suggest that the change in the \ac{CFL} thickness depends on the reduced volume of the \acp{RBC}.
It has been reported in the literature that cells that are close to the walls in a Poiseuille flow tend to migrate towards the center of the pipe due to their deformability~\cite{shi2012numerical}.
In contrast, a large volume fraction of \acp{RBC} may cause the cells to move closer to the walls due to collisions between the cells.
In the remainder of this study we investigate the deformations of \acp{RBC} and their trajectories and correlate these observations to the \ac{CFL} thickness.
The deformation of the cells are quantified by the rim angle (see below) and bending energy of each cell.
We distinguish tumbling-like motion, due to the external shear stresses, from tank-treading motion, where the membrane rotates around a steady shape.
The cell trajectories are quantified by the diffusion coefficient along the radial direction.
We fix $R=\SI{40}{\micro\meter}$, $\nabla p = \SI{1e-3}{\cmHHO\per\micro\meter}$ and $\Ht=0.45$ for clarity, but similar results were observed for the other conditions listed in \cref{fig:blood:velocity}.

\paragraph{Rim angle}

\begin{figure}
  \centering
  \includegraphics[width=\columnwidth]{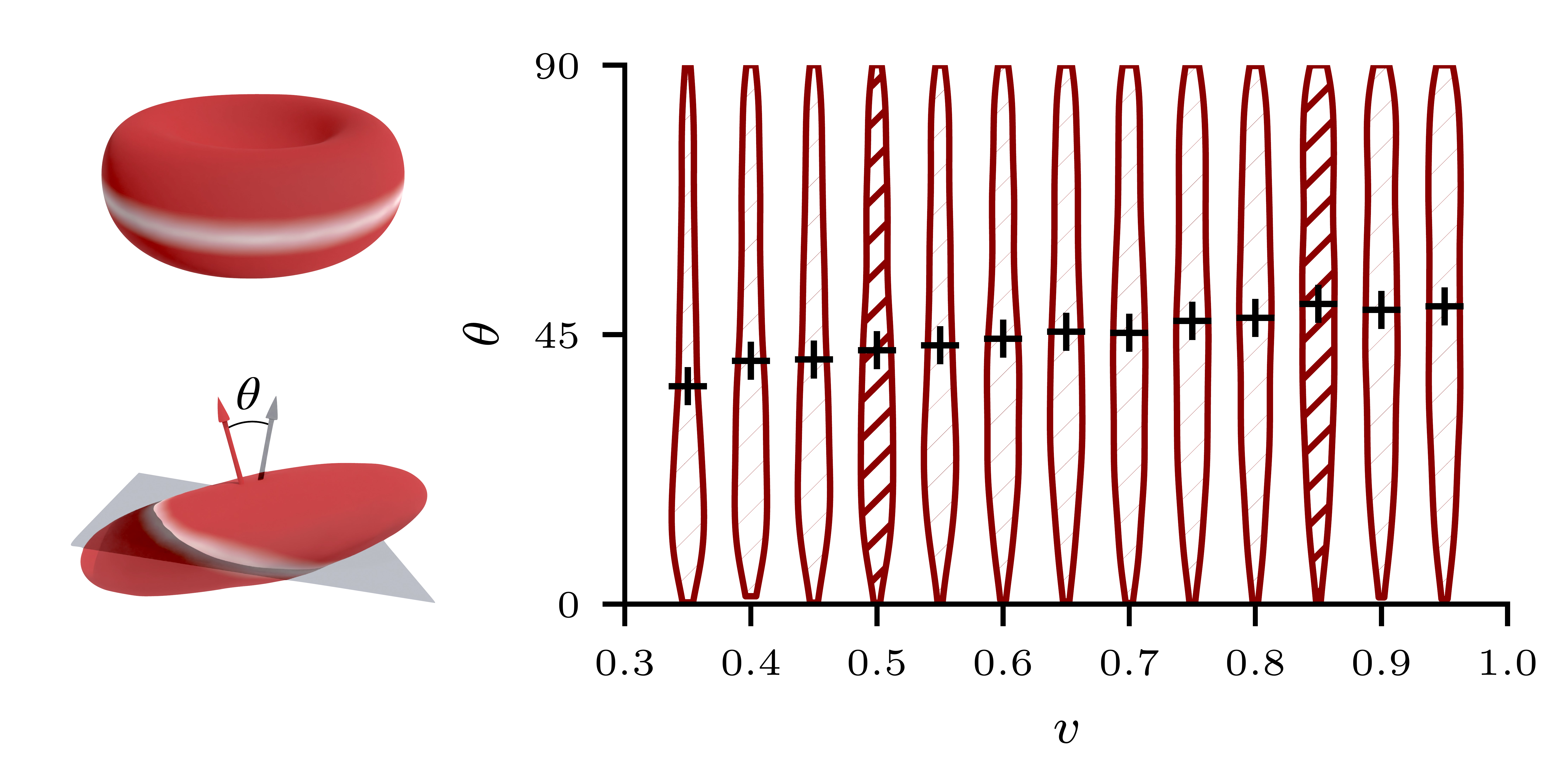}
  \caption{Top left: The rim (white) of a RBC at rest.
    Bottom left: The normal of the rim plane (dark arrow) and the normal of the cell plane (red arrow) form the rim angle $\theta$.
    Right: Violin plots illustrating the variation in rim angle distribution $\theta$ (in degrees) of flowing cells across different reduced volumes $v$.
    Crosses indicate the mean value of $\theta$.
    \label{fig:rim}
  }
\end{figure}

In a sufficiently high shear, the \ac{RBC} membrane rotates around the cytosol~\cite{fischer1980}.
In this situation, the region of the membrane that forms the rim at equilibrium differs from the principal disc of the deformed cell, unlike \acp{RBC} at rest.
We define the rim angle as the angle between the normal to the principal disc of the cell and the normal to the region corresponding to the rim of the stress-free state of the cell, see~\cref{fig:rim} and ref.~\cite{economides2020thesis}.
For a tumbling \ac{RBC}, this angle is typically low while for tank-treading \acp{RBC} the angle takes values in the whole range, $\theta \in [0, \pi/2]$.
\Cref{fig:rim} shows the distribution of rim angles $\theta$ of cells flowing in the tube for different values of $v$.
On average, the rim angles are smaller at low values of $v$.
Furthermore, the density of $\theta$ is more uniform at high values of $v$ than at low reduced volume values.
This suggests that \acp{RBC} with a large reduced volume tank-tread while their motion is closer to tumbling at lower values of $v$.

\paragraph{Bending energy}

\begin{figure}
  \centering
  \includegraphics[width=\columnwidth]{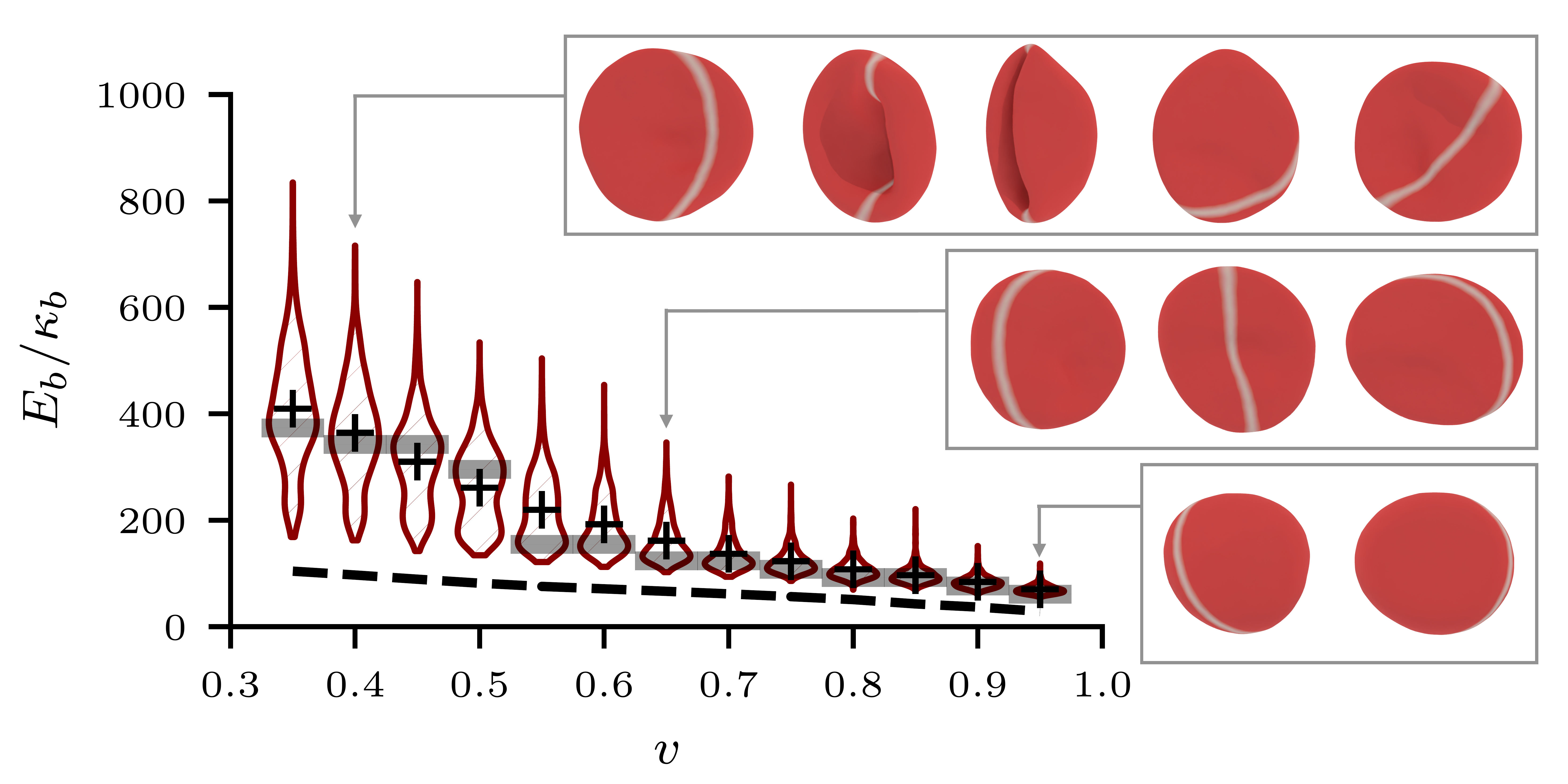}
  \caption{
    Left: Violin plots illustrating the variation in bending energy distribution of flowing cells across different reduced volumes $v$.
    Crosses indicate the mean value of the bending energy, shaded regions indicate the maximum density.
    The dashed line corresponds to the bending energy of the cells at rest.
    $\kappa_b$ is the bending modulus.
    Right: Time sequences of typical RBCs in the tube (time increasing from left to right). From top to bottom: $v=0.4$, $v=0.65$ and $v=0.95$. The white stripe represents the rim of the cell at rest (see \cref{fig:rim}).
    \label{fig:bending}
  }
\end{figure}

The deformations of the cells are characterized by the bending energy described by \cref{eq:energy:bending}.
This energy is high when the cell deforms (e.g. folds) and therefore characterizes complex dynamics and shape changes.
The bending energy distribution of the cells is shown on \cref{fig:bending} for different values of $v$.
The mean bending energy decreases when $v$ increases, consistent with the fact that the minimum bending energy is achieved for a spherical shape.
Nevertheless, we observe differences in the distributions of the bending energies.
For $v \geq 0.6$, the bending energies are concentrated close to a minimal value.
Instead, when $v < 0.6$, we observe a peak of bending energy well above the minimal values observed at each reduced volume.
These high values suggest that \acp{RBC} undergo large deformations compared to their resting shape.
In particular, we observe cells that are ``folding'' periodically when $v$ is relatively small (\cref{fig:bending}), as opposed to cells that have a large reduced volume.
These large deformations cause a large bending energy density in the bulk, or, equivalently, a large pressure compared to configurations with undeformed cells.
The larger deformations associated with low reduced volumes may thus contribute to reducing the \ac{CFL} thickness in this regime.
Furthermore, the deformation of \acp{RBC} is accompanied by an additional dissipation due to the membrane viscosity and the recirculation of the cytosol inside the cell~\cite{fischer1980}, which may further decrease the blood flux.

\paragraph{Tank-treading frequency}

\begin{figure}
  \centering
  \includegraphics[width=\columnwidth]{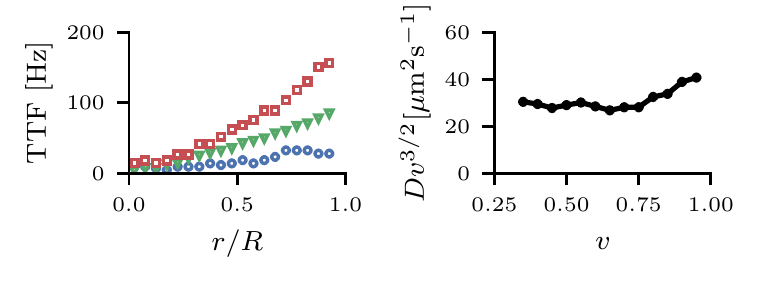}
  \caption{
    Left: TTF of the RBCs against their time averaged radial position, for $v=0.4$ (circles), $v=0.7$ (triangles) and $v=0.9$ (squares).
    Right: Radial diffusion $D$ of the RBCs against the reduced volume $v$.
    \label{fig:ttf_diff}
  }
\end{figure}

In linear shear flows, \acp{RBC} tank-tread when the shear rate is large enough~\cite{fischer2007,dupire2012}.
The \ac{TTF} is reported against the time averaged radial position of the cells on \cref{fig:ttf_diff} (see appendix for details).
For low values of $v$, most cells do not tank-tread hence the value estimated for the \ac{TTF} is low.
For larger values of $v$, the \ac{TTF} increases linearly with the radial position.
This increase is consistent with the nearly linear relationship between the \ac{TTF} and the shear rate for single cells~\cite{fischer2007}, and assuming that we have a nearly parabolic velocity profile along the radial direction of the tube, hence a linear increase of the shear rate with the radial position $r$.
Furthermore, we observe that the \ac{TTF} of cells is larger when $v$ increases.
This is also expected since spheres in shear flows rotate with a larger frequency than the \ac{TTF} of \acp{RBC}~\cite{fischer2007}.
We suggest that, for relatively thin cells, tank-treading is beneficial for cells to pass each other without deforming and to minimize the effect of their collisions, thus contributing to a lower value of $\delta$.
It was shown that tank-treading provides less hindrance to the motion of other particles compared to other motions like tumbling~\cite{kruger2013crossover}.
This may be an indication that tank-treading cells contribute to a lower value of $\delta$ compared to other motions, present notably at low values of $v$.

\paragraph{Radial diffusion}

\Acp{RBC} in blood flow undergo shear-induced diffusion, which is a result of the cell-cell interactions.
These interactions influence the radial distribution of the cells and the \ac{CFL} thickness~\cite{higgins2009statistical,srivastav2012shear}.
We thus measure the radial diffusion coefficient of the \acp{RBC} in the tube.
The diffusion coefficient is estimated from the mean squared displacement of the cells along the radial direction:
\begin{equation*}
  D = \langle \left(r(T)-r(0)\right)^2 \rangle / 2T,
\end{equation*}
where $r$ is the radial position of the cell's center of mass and $T$ is a time large enough to collect statistics but small enough to remain in the linear regime of the mean squared displacement against time.

The theory of shear-induced diffusion suggests that, at a given volume fraction, $D \propto \dot{\gamma} a^2$, where $\dot{\gamma}$ is the shear rate and $a$ is the size of the particle~\cite{rognon2021shear}.
Assuming that the shear rate is constant across the simulations and $a\propto v^{1/3}$, we expect $Dv^{3/2}$ to be approximately constant against $v$.
This quantity is reported on \cref{fig:ttf_diff}.
We observe that $Dv^{3/2}$ remains constant within a 30\% deviation, suggesting that increasing $v$ increases the shear-induced diffusion.
The deviations from the theory may come from the cells deformability, the definition of the cell radius from $v$ and the non-constant shear rate along the radial direction.
The shear-induced diffusion is higher for larger values of $v$ and may thus contribute to reducing the \ac{CFL} thickness for $v>0.65$~\cite{srivastav2012shear}, hence decreasing the transport efficiency of the cytosol.

\section*{Discussion}

The physiological volume of \acp{RBC}, under the assumptions made in this study, is the most efficient to transport oxygen through tubes that have the characteristics of arterioles.
When the reduced volume of the \acp{RBC} deviates from its physiological value, we identify two factors that decrease the transport efficiency of oxygen: the change of the \ac{CFL} thickness, and the dissipation due to the cytosol recirculation and the deformations of the cells.

The viscosity of plasma is smaller than that of whole blood, thus a larger \ac{CFL} thickness contributes to a higher flux of blood.
We find that the \ac{CFL} thickness is maximized at $v\approx 0.65$.
Below this value, cells fold, undergo large deformations and adopt a tumbling rather than tank-treading motion.
These deformations increase the effective thickness of the cells along the radial direction, thus reducing  the \ac{CFL} thickness.
In contrast, at larger values of $v$, cells do not deform and have more diffusive trajectories, due to collisions with their neighboring cells.
These collisions cause the cells to migrate perpendicularly to the flow direction, thus reducing the \ac{CFL} thickness.
At intermediate values of $v$, the cells tank-tread, which facilitate their motion relative to each other.
This is supported by the shear-thinning behavior of blood when the motion of \acp{RBC} transitions from tumbling to tank-treading~\cite{fedosov2011,forsyth2011}.
The non-spherical shape of \acp{RBC} limits the effect of collisions, and they form a maximal \ac{CFL} thickness.

\begin{figure}
  \centering
  \includegraphics[]{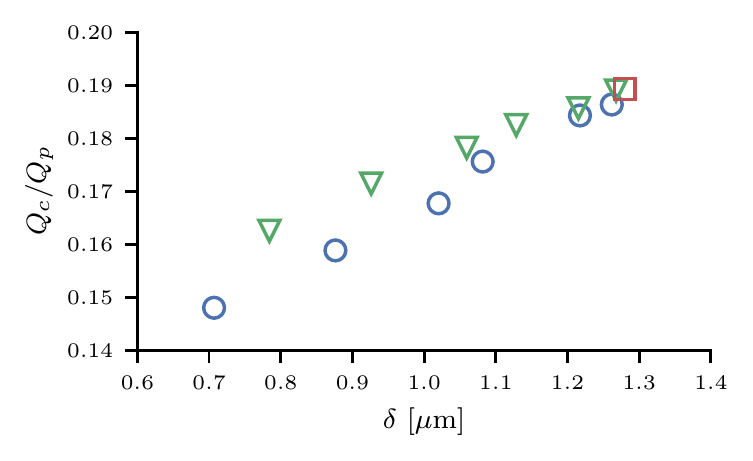}
  \caption{Cytosol flux against the CFL thickness.
    Circles correspond to cells with $v < 0.65$, triangles those with $v > 0.65$ and the square is for $v = 0.65$.
    \label{fig:cflQ}
  }
\end{figure}

The \ac{CFL} thickness alone does not explain the profile observed in \cref{fig:vQ}.
Indeed, for the same value of \ac{CFL} thickness, the cytosol flux is lower for the case with $v < 0.65$ than when $v > 0.65$ (\cref{fig:cflQ}).
We attribute this difference to the additional dissipation due to the large deformations occurring for cells with smaller reduced volumes, as membrane deformation was shown to be a dissipative process~\cite{tran1984}.
In addition, we expect that tank-treading cells dissipate more energy at lower reduced volumes, as shown for vesicles~\cite{kraus1996}.

In this study we tackled the following question: what is the reduced volume that corresponds to the highest flux of a given amount of hemoglobin?
We achieved this by adjusting the number of \acp{RBC}, assuming that the concentration of hemoglobin in the cytosol is the same in every situation.
However, we could also consider a hemoglobin concentration that varies with $v$, e.g. a higher concentration at low values of $v$ to model erythropoiesis~\cite{matoth1971postnatal}, but this approach would require to model the dependence of the cytosol viscosity with the hemoglobin concentration.
These different concentrations would change the volume of oxygen transported by the system, and the viscosity of the cytosol.
In this study we did not explore these effects due to the large computational cost of the simulations and made the approximation that the hemoglobin remains constant across all cases, which corresponds to the simplest model.
We leave the more complex scenario of varying the concentration as well as the number of \acp{RBC} to further studies.

The results of this study are valid for straight tubes, which is a simplistic approximation of arterioles.
The \ac{CFL} thickness is known to vary around bifurcations~\cite{rashidi2023red,bacher2018antimargination,zhou2021emergent,balogh2019cell} or curved geometries~\cite{spieker2021effects}.
However, the deformation of cells at low $v$ and the collisions of cells at larger values of $v$ might still contribute to reduce the \ac{CFL} thickness in those conditions compared to the physiological values, and we thus expect the same trend as in straight tubes.
Nevertheless, the optimal value of $v$ might slightly differ in those cases and further studies are required to provide a quantitative description of these variations.

Finally, we remark that we considered the advective part of oxygen transport in arterioles.
The diffusive transport of oxygen, not studied in this work, is responsible for transporting oxygen across the \ac{RBC} membranes.
This type of transport may be affected by the shape, and thus the reduced volume, of \acp{RBC}~\cite{lenard1974,uzoigwe2006,richardson2020}.

\section*{Conclusion}

We demonstrate, through high fidelity simulations, that \acp{RBC} with volumes similar to those observed in vivo maximize the efficiency of advective oxygen transport in straight tubes that have characteristics of arterioles, where the vascular resistance is maximal.
We qualitatively explain the variation of advective oxygen transport efficiency with respect to the reduced volume of the cells based on the \ac{CFL} thickness and the dissipation due to the cells deformability.
At low reduced volumes, cells deform and fold, thus occupy a larger effective volume on average.
This causes the \ac{CFL} thickness to decrease.
The deformations of cells induce additional dissipation due to the recirculation of the cytosol and the highly viscous membranes.
At large reduced volumes, cells collide and migrate in the directions perpendicular to the flow due to a high shear-induced diffusion.
These trajectories also contribute to reducing the \ac{CFL} thickness and thus the oxygen transport is lower.

The present findings provide valuable insights into the mechanisms of advective oxygen transport in the body and could potentially have significant implications for the advancement of therapies aimed at treating circulatory disorders.

\section*{Appendix}

\subsection*{Dissipative Particle Dynamics}

The \ac{DPD} method discretizes a fluid into $N$ particles with positions $\mathbf{r}_i$, velocities $\mathbf{v}_i$ and mass $m$, $i=1,2,\dots,N$.
The particles evolve according to the Newton's law of motion,
\begin{equation*}
  \dot{\mathbf{r}}_i = \mathbf{v}_i, \;\;\; \dot{\mathbf{v}}_i = \frac{1}{m} \sum \limits_{j=1}^{N} \mathbf{F}_{ij},
\end{equation*}
where $\mathbf{F}_{ij}$ are pairwise forces that vanish after a cutoff distance $r_c$.
These interactions are formed by three terms~\cite{hoogerbrugge1992,espanol1995},
\begin{equation*}
  \mathbf{F}_{ij} = a w(r_{ij}) \mathbf{e}_{ij}
  - \gamma \left(\mathbf{e}_{ij} \cdot \mathbf{v}_{ij}\right)  w_D(r_{ij}) \mathbf{e}_{ij}
  + \sigma \xi_{ij} w_R(r_{ij}) \mathbf{e}_{ij},
\end{equation*}
where $\mathbf{v}_{ij} = \mathbf{v}_i - \mathbf{v}_j$, $\mathbf{r}_{ij} = \mathbf{r}_i - \mathbf{r}_j$, $r_{ij}$ = $\|\mathbf{r}_{ij}\|$ and
$\mathbf{e}_{ij} = \mathbf{r}_{ij} / r_{ij}$.
The coefficients $a$, $\gamma$ and $\sigma$ are the conservative, dissipative and random force magnitudes, respectively.
Furthermore, we use the standard \ac{DPD} conservative kernel
\begin{equation*}
  w(r) =
  \begin{cases}
    1 - r/r_c, & r < r_c,\\
    0, & \text{otherwise}.
  \end{cases}
\end{equation*}
We set $w_D = w^{1/4}$.
$w_R$ satisfies the fluctuation-dissipation theorem, $\sigma^2 = 2 \gamma k_BT$ and $w_D = w_R^2$~\cite{espanol1995}, where $k_BT$ is the temperature of the system in energy units.

\subsection*{Parameters of the model}

The values of the \ac{RBC} parameters are listed in \cref{tab:params}.
They correspond to the mean value of the posterior distribution found in ref~\cite{amoudruz2023stress}.

\begin{table}[h]
  \centering
  \begin{tabular}{l|l}
    Parameter & Value \\
    \hline
    $\kappa_b$ & $\SI{2.10e-19}{\joule}$ \\
    $\mu$ & $\SI{4.99}{\micro\newton\per\meter}$ \\
    $K_\alpha$ & $\SI{4.99}{\micro\newton\per\meter}$ \\
    $a_3$ & $-2$ \\
    $a_4$ & $8$ \\
    $b_1$ & $0.7$ \\
    $b_2$ & $1.84$ \\
    $\eta_m$ & $\SI{0.42e-6}{\pascal\second\meter}$ \\
    $A_0$ & $\SI{135}{\micro\meter^2}$ \\
    $V_0$ & varying \\ 
    $k_A$ & $\SI{0.5}{\joule\per\meter^2}$ \\
    $k_V$ & $\SI{7.23e5}{\joule\per\meter^3}$
  \end{tabular}
  \caption{Parameters of the RBC model.
    \label{tab:params}
  }
\end{table}

\subsection*{Computation of the cell-free layer thickness}

The cell-free layer thickness $\delta$ is computed from the vertices of the \ac{RBC} membranes.
For a given time snapshot, we compute the maximum radial position of vertices over bins placed at regular intervals along the flow direction.
The bins have a size of $\SI{1}{\micro\meter}$.
The value of $\delta$ at a given time is then the difference between the radius of the pipe and the average of this quantity over all bins.
The values reported in this work are the average of the cell-free layer thickness over time, after equilibration of the flow.




\subsection*{Computation of the tank-treading frequency}

The \ac{TTF} is estimated by computing the Fourier transform of the rim angle time series of each cell and selecting the frequency of the largest mode, divided by 2 since the rim angle of a tank-treading cell undergoes 2 revolutions per tank-treading revolution.
Note that we exclude cells having an average rim angle $\bar{\theta}$ larger than $\pi/4$, and the tumbling cells, which we characterize by $\bar{\theta} < \pi/8$.

\section*{Author contributions}

LA, AE and PK designed research. LA performed research. LA and AE analyzed data. LA, AE and PK wrote the manuscript.

\section*{Declaration of interests}

The authors declare no competing interests.

\section*{Acknowledgments}

We would like to thank Xin Bian for the idea of using the rim angle to perform the analysis.
We acknowledge the computational resources granted by the Swiss National Supercomputing Center (CSCS) under the project ID “s1160”.

\bibliographystyle{plain}
\bibliography{bibliography}

\end{document}